# Views about ChatGPT: Are human decision making and human learning necessary?

Eiji Yamamura, Fumio Ohtake

# Abstract

Using individual-level survey data from 2024, this study investigated how respondent characteristics are associated with a subjective view of generative artificial intelligence (GAI). We asked 14 questions concerning respondents' view about GAI, such as general view, faulty GAI, autonomous GEI, GAI replacing humans, and importance of human learning. Regression analysis based on the ordered logit model revealed that: (1) In some cases, the results of smartphone and computer usage times differed. Smartphone usage time was negatively correlated with the importance of human learning, whereas computer usage was not negatively correlated. (2) Managers and ordinary businesspeople have positive views of GAI. However, managers do not show a positive view about GAI being responsible for human decision making. (3) Teachers generally have a negative view about GAI replacing humans and no need of learning. They do not have negative views about GAI producing documents unless GAI is faulty. (4) Medical industry workers positively view GAI if it operates following their direction. However, they do not agree with the view that GAI replaces humans, and that human learning is unnecessary. (5) Females are less likely than men to have a positive view of GAI. In summary, views about GAI vary widely by the individual's characteristics and condition of GAI, and by the question set.

Key words: Subjective view, ChatGPT, Teacher, Managers, Smartphone usage time, Computer usage time, Gender differences.

# 1 Introduction

As a type of generative pretrained transformer (GPT), ChatGPT was developed by OpenAI and released in November 2022. Thus, the general public could use large language models (LLM) for free; notably, ChatGPT had over one million users within five days from its release [1]. ChatGPT can create humanlike text, engage in conversations, and answer questions with a human-like understanding. ChatGPT can also create an image of, say, "economics" if the user asks ChatGPT to illustrate it. Thus, ChatGPT can create both text and images following human directions.

Demonstrating the high interest in it, search volumes for ChatGPT on Google surged constantly for six months since its launch [2]. However, generative artificial intelligence (GAI) covers wider fields than LLM. For instance, Hollywood's Union for Actors went on strike mainly because actors rejected a proposal from companies for using AI technology to scan their faces and bodies [3]. By using machine learning algorithms based on various types of online information, GAI is expected to play a critical role in creation things, which has traditionally been conducted by humans. Indeed, about ten years ago, much attention was drawn to the list of occupations that would disappear because of AI's emergence [4], and this list has continued to grow because of unexpected technological developments.

Still, predicting the impact of GAI, which has a broader range of functions and meanings than LLM, on occupation and learning itself is actually difficult. Anyone can easily use GAI owing to recent advancements in this field. This can trigger social transformation, and GAI is now considered as a "game-changer that society and industry need to be ready for" [5]. GAI has appeared as a coauthor in research papers, leading researchers to discuss the issue of GAI in the academic world [6]. It also influences journalism and media education [7]. To evaluate the quality of GAI writing, a study compared abstracts written by humans and artificial general intelligence, and found that artificial general intelligence creates a more readable abstract, but its overall quality is lower [8]. Crucially, GAI is anticipated to reduce the inequality in writing skills between more- and less-educated people [9]. Furthermore, human decision-making may be replaced by GAI [10]. A survey study investigated the subjective views of GAI by comparing those who used GAI with others [11].

In the business world, GAI is useful for generating human-like documents and promptly translating text across many languages. Further, for professional writing tasks, GAI substantially increased productivity by reducing the writing time by 40% and increasing output quality rose by 18% [9]. GAI also helps reduce unproductive and routine work, such as producing routine documents and deadlocked meetings in the business world, and communicating with foreigners without the bearing cost of learning foreign languages. For instance, the "academic performance" of GAI exhibited high performance in an experiment in a management course. Moreover, its answers were correct and the explanations were excellent for basic operations management and process analysis questions [12].

GAI also influences education and learning methods. In educational management, GAI can improve teaching and educational resource allocation. GAI also provides students with more personalized learning experiences tailored to their learning abilities [13]. For instance, some teachers and students agreed that the use of GAI in schoolwork should be acknowledged [14]. Furthermore, they believe that GAI will improve the competitiveness of students who are nonnative English speakers. Students also note that in their future jobs, they will be able to outsource mundane tasks to ChatGPT, allowing them to focus on substantive and creative work. GAI is expected to significantly affect various fields such as cyber security, customer support, software development [15], and medical examinations [8,16].

Overall, GAI is anticipated to be beneficial in real-world situations in various fields. GAI is likely to spread rapidly throughout society and influence human life, including ways of working and learning in education. This naturally gives individuals a great incentive to adjust to changing lifestyles, even if they are reluctant to do so.

However, several concerns remain about GAI. For example, misinformation is often included in GAI while creating content. The drastic diffusion of GAI has rapidly produced large amounts of text, further catalyzing the spread of misinformation. This can be considered as "GAI-driven infodemic" [17,18]. However, it is difficult to distinguish misinformation partly because humans miss the randomly inserted misinformation by GAI because they read it in a naturally readable text. GAI does not provide references or sources for the content used, and thus, the cost of discovering misinformation increases. In addition, GAI may randomly add sentences to documents at its discretion, even if their addition is not directed by a human user. Researchers have recently drawn attention towards AI hallucinations [19,20]. For example, after researchers ordered ChatGPT to provide studies related to a certain topic, it suggested several academic papers [20]. However, none of them existed and their PubMed IDs were from different unrelated studies. Clearly, the credibility and authenticity of GAI creating documents are not be taken at face value. Inevitably, humans should evaluate the costs and benefits of using GAI, even when simply producing documents.

Despite these concerns, with far lower costs than humans, GAI may exhibit almost equivalent performance and quality for various tasks in the near future. GAI is highly likely to replace humans in various fields and situations. Inevitably, society has reconsidered the significance and meaning of human roles and learning. This study deals not only with the impact of current LLMs, but also that of GAI with broader functions. Specifically, this study explores individuals' subjective views about GAI not only limited to producing documents but also related to the necessity of the human role and human learning. Furthermore, we compare views across works, genders, ages, and experiences with computers and smartphones.

## 2 Data

### 1.1. Data collection

From 2016 to the present, we have constructed panel data by conducting a long-term panel survey of the Survey on living environment during childhood, and Current Life Attitudes. As part of this survey, items related to COVID 19 pandemic were inserted in from the 2020 survey with a focus on COVID-19. So, apart from main title of the overall project, the specific title "A Study on the Influence of the New Type of Coronavirus Infection on the Lifestyle Consciousness" is added in surveys after 2020. The data used in this survey was collected in February 2024, when COVID-19 had almost ended. In addition to the previous questions, in 2024 survey, questions related to the impact of GAI on human life were newly inserted. This study used only data collected in 2024 survey.

To conduct the survey, we selected the Nikkei Research Company (NRC) because it has extensive experience in conducting academic surveys. The survey was conducted until a sufficiently sized sample was obtained. To collect data used in this study, NRC sent questionnaires to the participants via the Internet between February 1 to 9, 2024.

Consequently, 5,133 respondents participated. Some participants did not respond to some questions. Therefore, they were excluded from the sample used for estimation. The sample size for the analysis was reduced to 3,433 observations.

The questionnaires included questions about participant characteristics such as age, gender, household income, residential prefecture, and job status. In addition, we examined smartphone and computer usage times. As dependent variables, 14 questions asked the participants about their subjective views of GAI. Table 1 provides the definitions of the key variables, their mean values, and standard errors. Concerning views on GAI, the respondents were requested to answer the questions in the order presented in Table 1.

Eight questions were related to GAI-producing documents. The remaining six questions were related to the more intrinsic value of human existence under the condition that GAI performs equivalent to that of humans. That is, questions were related to the extent to which GAI should replace humans in the labor market and whether human learning is necessary.

Overall, the 14 views about GAI were classified into six categories. (1) Basic views: *Typical* and *Intellectual*, (2) faulty GAI: *Typical Faulty* and *Intellectual Faulty*, (3) autonomous GAI: *Typical Autonomous* and *Intellectual Autonomous,* (4) GAI replacing human: *Self Typical* and *Self Intellectual,* 5) Role of GAI: *Confidential, Human decision* and *GAI decision,* and (6) Need for human learning: *No Writing, No Foreign* and *No Learning*. Views (1) to (4) were limited to specific tasks of GAI under various settings. Views (5) and (6) were included to consider the intrinsic value of humans.

In Table 1, both smartphone and computer usage times ranged between 0.5 and 15, while their maximum values were 24. One may reasonably assume that the minimum value is greater than 0. However, 24 hour usage seems exaggerated. Indeed, in this study's survey, observations for usage times exceeding 15 were less than 1%. One may consider usage above 15 hours to be outliers or untrustworthy. However, there seem to be various ways to interpret question regarding smartphone or computer usage times. For instance, besides active using hours, the hours when these devices were switched on can be considered as usage time for some persons. Therefore, the hours over 15 are replaced to 15.

**Table 1: Definitions of key variables and their descriptive statistics.**

| Variables | Definition | Mean | S.D. | Min. | Max. |
|---|---|---|---|---|---|
| *Dependent Variables* | By entering information, you can now allow a generative AI to produce natural, human-written documents on your behalf. Please choose your opinion on this from five choices where larger values show a positive view, Respondent's choices: 1, 2, 3, 4, 5. | | | | |
| *Typical* | Letting a generative AI create typical documents. | 3.56 | 1.01 | 1 | 5 |
| *Intellectual* | Letting a generative AI produce intelligent documents. | 3.08 | 1.08 | 1 | 5 |
| *Typical Faulty* | When a generative AI produces a typical document, it does not care that misinformation will be mixed in. | 2.48 | 1.10 | 1 | 5 |
| *Intellectual Faulty* | When a generative AI produces intelligent documents, it does not care that misinformation is mixed in. | 2.44 | 1.10 | 1 | 5 |
| *Typical Autonomous* | When a typical document is produced, it does not care that the generative AI may mix in partially undirected content at its discretion. | 2.61 | 1.07 | 1 | 5 |
| *Intellectual Autonomous* | When an intelligent document is produced, it does not care that the generative AI may mix in partially undirected content at its discretion. | 2.59 | 1.08 | 1 | 5 |
| *Self Typical* | Letting the generative AI produce typical | 3.35 | 1.06 | 1 | 5 |
| | documents instead of yourself. | | | | |
| *Self Intellectual* | Letting the generative AI produce intelligent documents instead of yourself. | 3.03 | 1.09 | 1 | 5 |
| *Confidential* | Generative AI can also handle the creation of confidential documents, eliminating the need for human involvement. | 2.26 | 1.05 | 1 | 5 |
| *Human decision* | If generative AI is able to perform various tasks on behalf of humans, humans only need to make final decisions. | 2.67 | 1.06 | 1 | 5 |
| *GAI decision* | As generative AI improves, it is better for generative AI to make decisions on management and other matters, in addition to various practical tasks. | 2.53 | 1.04 | 1 | 5 |
| *No Writing* | Generative AI has advanced document creation capabilities, so there is no need for humans to develop writing skills. | 2.28 | 1.07 | 1 | 5 |
| *No Foreign* | With generative AI translation, there is no need for humans to learn a foreign language. | 2.56 | 1.10 | 1 | 5 |
| *No Learning* | As generative AI develops and becomes as capable as humans, there is no need for humans to learn and improve their capabilities. | 2.19 | 1.07 | 1 | 5 |
| *Independent Variables* | | | | | |
| UNIV | It is one if a respondent graduated from university, and otherwise 0. | 0.55 | 0.49 | 0 | 1 |
| SMART | Answer to "How much time a day do you spend using smartphone?" Time spent using smartphone (hours) | 2.18 | 2.27 | 0.5 | 15 |
| COMPUTER | Answer to "How much time a day do you spend using computer?" Time spent using computer (hours) | 3.79 | 3.21 | 0.5 | 15 |
| INCOME | Household income (million yens) | 660 | 475 | 50 | 2300 |
| FEMALE | It is one if a respondent is female, and otherwise 0. | 0.46 | 0.49 | 0 | 1 |
| AGE | Respondents' ages | 47.6 | 13.2 | 19 | 73 |
| TOP MANAGE | It is one if a respondent is an executive or a manager in a company, and otherwise 0. | 0.10 | 0.29 | 0 | 1 |
| ORDINARY | It is one if a respondent is an ordinary company employee, and otherwise 0. | 0.30 | 0.46 | 0 | 1 |
| TEACHER | It is one if a respondent is a teacher, and otherwise 0. | 0.02 | 0.12 | 0 | 1 |
| SPECIAL | It is one if a respondent is a lawyer, accountant, and in other licensed occupations, and otherwise 0. | 0.03 | 0.16 | 0 | 1 |
| MEDICAL | It is one if a respondent is a medical worker, including medical doctor and nurse, and otherwise 0. | 0.01 | 0.08 | 0 | 1 |

Figs 1-6 illustrate histograms of views about GAI. Fig 1 (1) and (2) represent *Typical* and *Intellectual,* respectively. As for *Typical*, almost half of individuals chose 4 or 5, showing a positive view about GAI producing typical documents. However, the percentage of those with a positive view declined to 30% for *Intellectual*. Hence, approximately 20% individuals consider that GAI can produce typical documents, but not intellectual ones.

**Fig 1. General views about AI.**

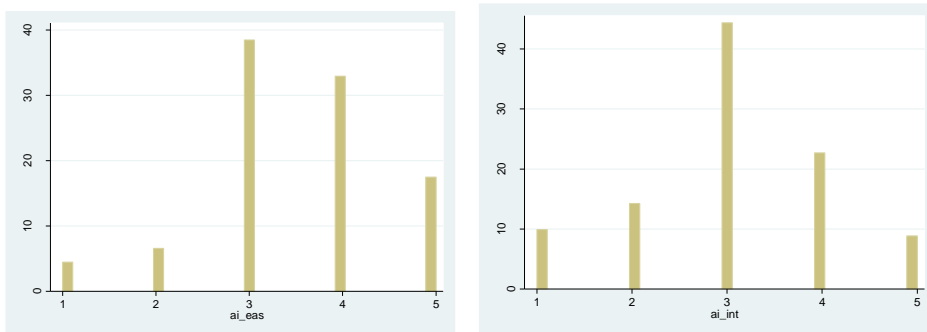

(1) Typlical    (2)  Intellectual

**Fig 2. Views about faulty AI.**

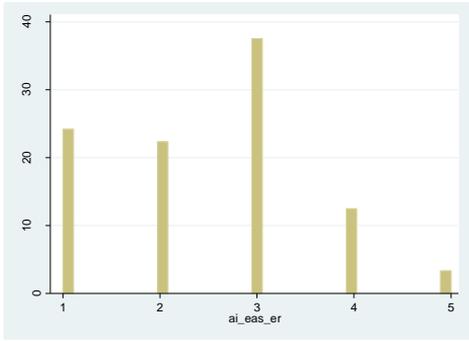

(1) *Typical Faulty*

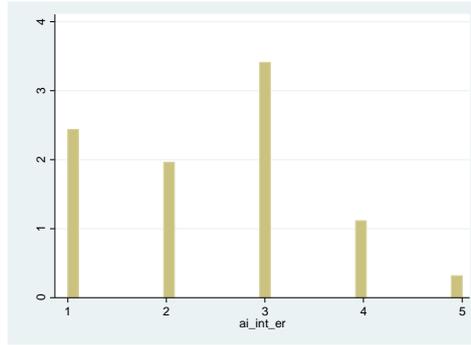

(2) *Intellectual Faulty*

Fig 3. Views about autonomous AI.

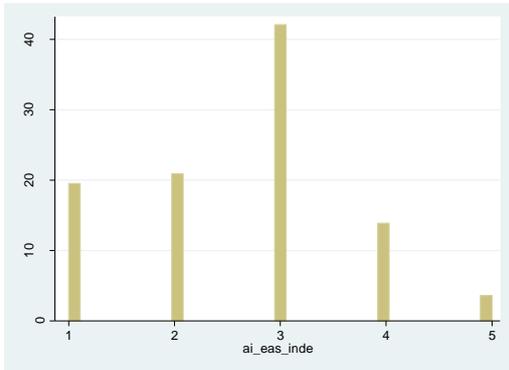

(1) *Typical Autonomous*

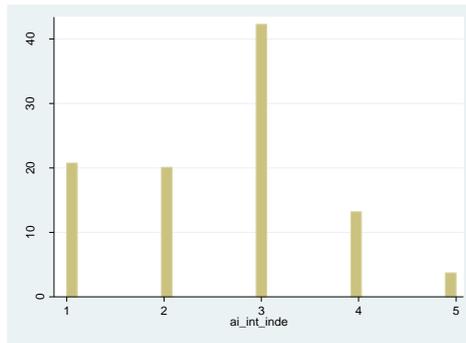

(2) *Intellectual Autonomous*

Fig 4. Views about AI replacing humans.

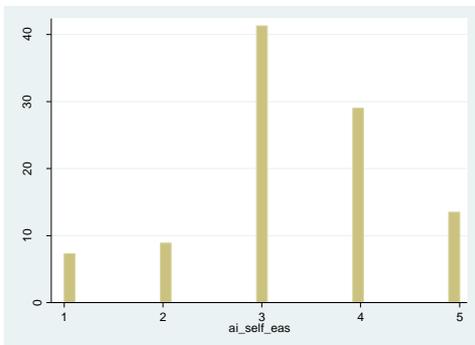

(1) *Self Typical*

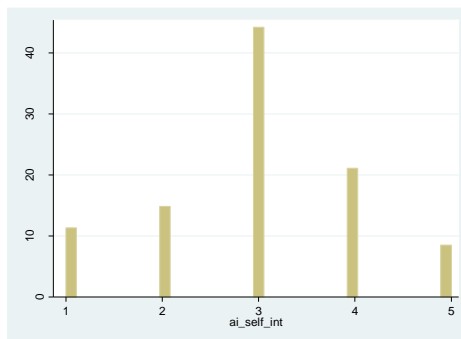

(2) *Self Intellectual*

Fig 5. Views about the role of AI.

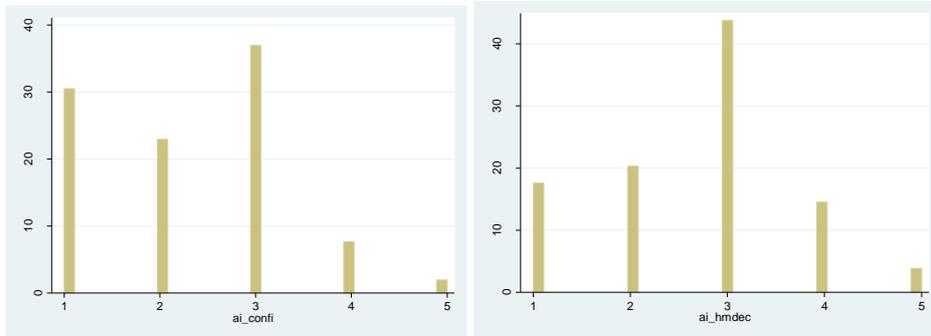

(1) Confidential          (2) Human decision

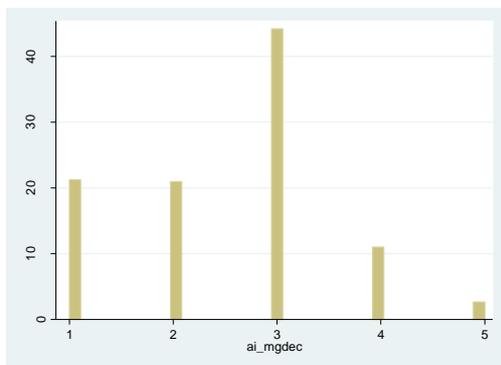

(3) GAI decision

Fig 6. Views about the importance of human learning.

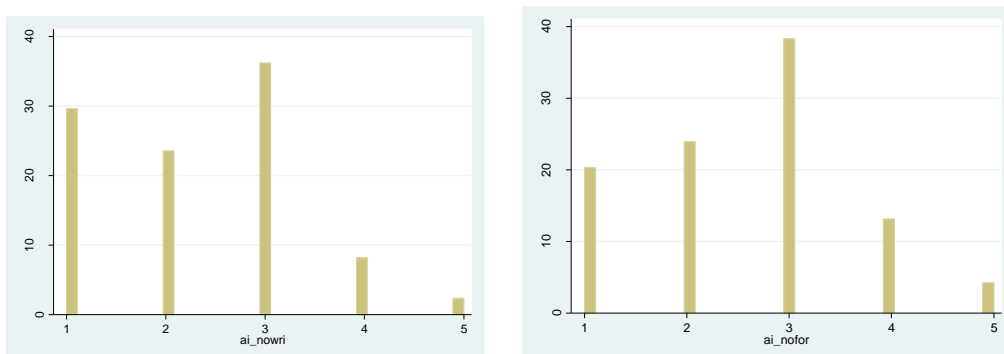

(1) No writing          (2) No foreign

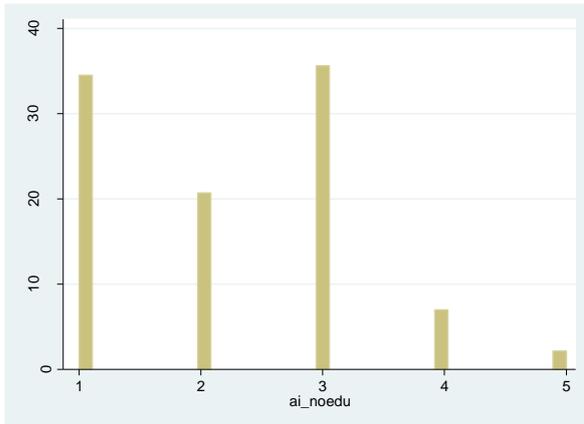

*(3)No learning*

Fig 2 demonstrates that GAI produced documents with misinformation. The proportion of positive views drastically reduced to approximately 15% from 30% in Figure 1, regardless of the type of document. That is, over half of the individuals with a positive view in Fig 1 care that misinformation will be included in the documents. Similar to Fig 2, Fig 3 indicates that the percentage of positive views on autonomous GAI was approximately 15%. Over half of the individuals with a positive view in Fig 1 care that AI may mix in partially undirected content at its discretion. Fig 4 (1) and (2) are similar to Fig 1(1) and (2).

Fig 5 (1)–(3) indicate the role of GAI. Fig 6 (1)-(3) illustrate human learning. In most cases, only approximately 10% of respondents had the positive views. Thus, the majority of respondents did not support the view that GAI reduces the role of humans in work or the need for learning.

## 2.2 Ethical considerations

Before starting the survey, participants are informed of the purpose and content of the survey and their consent is obtained. Even after starting the survey, participants had the right to quit at any time. As conditions for survey participants, those who are over ages of 18 who are considered an adult by Japanese law when we conducted the survey. Therefore, there is no need of obtaining consent from parents or guardians.

This study's survey design of was conducted with ex ante approval from the Ethics Committee of the Graduate School of Economics, Osaka University (approval no.: R51127). We have obtained the approval in November 2023 before the survey in Feb 2024. Data collection was performed in accordance with relevant guidelines and regulations. This study did not involve experimental manipulation, there were no foreseeable risks involved, the questionnaire used for this research was anonymous.

As is explained in the section 2.1, the data used in this study are part of a long-term panel data of lifestyle attitudes that includes survey of the new coronavirus infections. The application documents for ethics review clearly state the purpose of the study as "to accumulate basic knowledge for more desirable policy formulation based on changes in social awareness after COVID 19 pandemic." The data related to the generated AI used in this study were collected for this purpose. Therefore, the intent of this study is consistent with the project title "Survey on living environment during childhood, and Current Life Attitudes" that is indicated as subtitle of the survey in ethics approval although the main tile in the ethics approval is "A Study on the Influence of the New Type of Coronavirus Infection on the Lifestyle Consciousness".

## 3  Method

Based on their definitions, the dependent variables have discrete values between 1 to 5. Therefore, an ordered logit model is preferred for the estimation. For instance, the estimation results can be obtained for the probability of choosing 5, indicating that the respondent chose a very positive view of the question. The results for the remaining chosen values can be similarly obtained. For simplicity, we only report the results when the respondents chose five (very positive) or one (very negative). We can calculate the marginal effect for the dependent variables in each outcome of participant views. The independent variables include both continuous variables such as SMART, COMPUTER,

INCOME, and AGE, and binary dummy variables such as FEMALE, UNIV, TOP MANAGE, ORDINARY, and TEACHER. The effects of these two sets of variables may not be comparable because their scales differ. In economics, elasticity is commonly used as a standardized effect to compare variables with different scales. Elasticity is the percentage change in the dependent variable when the independent variable increases by 1 percent. TFor instance, consider the probability that the respondent chooses 5. Then, the elasticity is calculated as follows:

$$\frac{\bar{x}}{\bar{y}}\frac{\partial y}{\partial x}$$

$\frac{\partial y}{\partial x}$ is marginal effect of x for prob (y=5). $\bar{x}$ and $\bar{y}$ are the mean values of x and y, respectively. To control the scale effect, $\frac{\bar{x}}{\bar{y}}$ is multiplied to the marginal effect. Furthermore, the z values are obtained using the delta method. We used Stata 14 to estimate the ordered logit model by "Ologit" and then elasticity by "margins,eyex" added directly below the command of "Ologit."

The model assesses the association between the respondents' characteristics and their views on GAI. The estimation function takes the following form:

$$Y_i = \alpha_0 + \alpha_1 UNIV_i + \alpha_2 SMART_i + \alpha_3 COMPUTER_i + \alpha_4 INCOM_i + \alpha_5 FEMALE_i$$
$$+ \alpha_6 AGE_i + \alpha_7 TOP\ MANAGE_i + \alpha_8 ORDINARY_i + \alpha_9 TEACHER_i$$
$$+ \alpha_{10} SPECIAl_i + \alpha_{11} MEDICAL_i + X_i A_i + e_i$$

$Y_i$ is the dependent variable. As we have 14 views, $Y_i$ is replaced based on these views. Hence, 14 estimations were conducted. α denotes the elasticity of the variables and has five different values in a estimation because there are probability that $Y_i$ has 1, 2, 3, 4, and 5. *i* represents an individual. X is the vector of the control variables and A is the vector of their elasticity. X represents the residential prefecture and is a dummy variable for medical workers. The results for medical workers do not show statistical significance, and thus, have not been reported. $e_i$ is an error term.

The key independent variables are SMART, COMPUTER, TOP MANAGE, ORDINARY, and TEACHER. Experience with smartphones and computers may lead individuals to consider evaluating the strengths and weaknesses of GAI more correctly. To compare different works, TOP MANAGE, ORDINARY, TEACHER, SPECIAL, and MEDICAL are incorporated. GAI is considered a game-changer in the business world. Therefore, we investigate how businesspeople evaluate and expect GAI in their businesses. Furthermore, in education, the appropriate use of GAI has drawn great attention; hence, TEACHER is included. In SPECIAL, for instance, accountants may be replaced with AI. That is, GAI might be a substitute for human workers in SPECIAL cases [4]. These special workers lose their jobs, and therefore, are likely to negatively view GAI. Research suggests that GAI is being practically used in medicine and medical practice.

Even before the emergence of ChatGPT, the use of electronic records in healthcare was much ahead of other industries [21,22]; moreover, AI has been used for image diagnostics [23–25]. GAI is considered complementary to humans in medical areas; thus, workers in the medical sector are predicted to be positive about GAI. Finally, research shows that females tend to be more risk-averse than male. A particular problem with GAI is that it can mix misinformation in its creations, which is considered a type of risk. Hence, FEMALE is incorporated to explore how females consider this problem.

# 4 Results

Tables 2-7 report probability of having very positive view (choosing 5) in the upper part and that of having very negative view (choosing 1) in the lower part. The signs of elasticity are opposite between them. Basically, the results of positive view are consistent with those of negative one. We interpret the results of positive view.

## 4.1 Results on general views

**Table 2. Dependent variable: General views about GAI (ordered logit model). Elasticity (standardized effect).**

|  | (1) | (2) |
|---|---|---|

|  | *Typical* | *Intellectual* |
|---|---|---|
| **Prob (=5)** | | |
| UNIV | −0.002 | −0.008 |
|  | (−0.01) | (−0.28) |
| SMART | 0.021 | 0.053* |
|  | (0.74) | (1.73) |
| COMPUTER | 0.156*** | 0.094** |
|  | (6.29) | (2.52) |
| INCOME | 0.225*** | 0.137*** |
|  | (5.54) | (2.69) |
| FEMALE | 0.009 | −0.119*** |
|  | (0.40) | (−3.92) |
| AGE | −0.426*** | −0.646*** |
|  | (−4.22) | (−3.90) |
| TOP MANAGE | 0.004 | 0.021** |
|  | (0.52) | (2.41) |
| ORDINARY | −0.001 | 0.029 |
|  | (−0.07) | (1.36) |
| TEACHER | −0.001 | −0.007 |
|  | (−0.42) | (−1.31) |
| SPECIAL | −0.001 | −0.003 |
|  | (−0.41) | (−063) |
| MEDICAL | 0.005** | 0.004** |
|  | (2.46) | (2.18) |
| **Prob (=1)** | | |
| UNIV | 0.002 | 0.008 |
|  | (0.01) | (0.28) |
| SMART | −0.025 | −0.053* |
|  | (−0.74) | (−1.71) |
| COMPUTER | −0.188*** | −0.094** |
|  | (−6.13) | (−2.48) |
| INCOME | −0.272*** | −0.138*** |
|  | (−5.38) | (−2.65) |
| FEMALE | −0.010 | 0.113*** |
|  | (−0.40) | (4.06) |
| AGE | 0.493*** | 0.632*** |
|  | (4.26) | (3.94) |
| TOP MANAGE | −0.004 | −0.023** |
|  | (−0.52) | (−2.33) |
| ORDINARY | 0.001 | −0.029 |
|  | (0.07) | (−1.35) |
| TEACHER | 0.001 | 0.007 |
|  | (0.43) | (1.49) |
| SPECIAL | 0.011 | 0.003 |
|  | (1.64) | (0.65) |
| MEDICAL | −0.006** | −0.004** |
|  | (−2.04) | (−2.02) |
| Pseudo $R^2$ | 0.01 | 0.02 |
| Observations | 3,433 | 3,433 |

Note: ***, **, and * denote statistical significance at the 1%, 5%, and 10% levels,

respectively. The numbers within parentheses are the z-values calculated using robust standard errors at the residential prefecture level. Numbers without parentheses indicate elasticity, which is the standardized effect. Various control variables are included in all columns, such as residential prefecture and a dummy variable for marital status. Furthermore, the probabilities that the dependent variables are 2, 3, and 4 have been calculated but are not reported here due to save the space.

SMART and COMPUTER show positive signs and statistical significance, except for column (1) of SMART. Overall, the computer and smartphone usage times lead to a positive view of GAI. A significantly positive sign of INCOME indicates that higher-income people are more likely to positively view GAI.

By contrast, AGE has a negative sign in columns (1) and (2). Older adults may be unable to catch up with technological changes. FEMALE also shows a negative sign, although statistical significance is only observed in column (2). Interestingly, using GAI to produce typical documents is less likely to result in errors than when producing intellectual documents. This is consistent with female's risk aversion tendency [26–29].

Except for TOP MANAGE in column (2), TOP MANAGE, ORDINARY, TEACHER, and SPECIAL do not show any statistical significance. Consistent with this prediction, MEDICAL has a positive sign, which might reflect the fact that electronic medical records have been widely used in hospitals, playing a critical role in real situations [21,22].

## 4.2 Results on views about faulty AI

**Table 3. Dependent variable: Views about faulty AI (ordered logit model).** Elasticity (standardized effect).

|  | (1) *Typical Faulty* | (2) *Intellectual Faulty* |
|---|---|---|
| **Prob (=5)** | | |
| UNIV | −0.032 | −0.078** |
|  | (−1.01) | (−2.34) |
| SMART | 0.095*** | 0.090*** |
|  | (3.71) | (4.30) |
| COMPUTER | −0.065* | −0.081** |
|  | (−1.88) | (−1.99) |
| INCOME | −0.033 | 0.014 |
|  | (−0.56) | (0.27) |
| FEMALE | −0.167*** | −0.180*** |
|  | (−5.42) | (−5.13) |
| AGE | −0.863*** | −0.905*** |
|  | (−6.25) | (−7.13) |
| TOP MANAGE | 0.030*** | 0.025** |
|  | (3.16) | (2.28) |
| ORDINARY | 0.106*** | 0.119*** |
|  | (7.06) | (7.48) |
| TEACHER | −0.007** | −0.009*** |
|  | (−2.20) | (−2.88) |
| SPECIAL | −0.001 | 0.007 |
|  | (−0.30) | (1.28) |
| MEDICAL | 0.001 | −0.001 |
|  | (0.37) | (−0.36) |
| **Prob (=1)** | | |
| UNIV | 0.024 | 0.058** |
|  | (1.02) | (2.38) |
| SMART | −0.078*** | −0.071** |
|  | (−3.66) | (−4.24) |
| COMPUTER | 0.050* | 0.060** |
|  | (1.91) | (2.02) |
| INCOME | 0.025 | −0.011 |
|  | (0.56) | (−0.27) |
| FEMALE | 0.122*** | 0.127*** |
|  | (5.78) | (5.42) |
| AGE | 0.661*** | 0.671*** |
|  | (6.26) | (7.15) |
| TOP MANAGE | −0.025*** | −0.020** |
|  | (−2.98) | (−2.16) |
| ORDINARY | −0.090 | −0.100*** |
| | (−6.62) | (−7.13) |
| TEACHER | 0.004*** | 0.005*** |
| | (2.70) | (4.01) |
| SPECIAL | 0.001 | −0.005 |
| | (0.31) | (−1.21) |
| MEDICAL | −0.001 | 0.001 |
| | (−0.35) | (0.38) |
| Pseudo $R^2$ | 0.02 | 0.03 |
| Observations | 3,433 | 3,433 |

Note: ***, **, and * denote statistical significance at the 1%, 5%, and 10% levels, respectively. The numbers within parentheses are the z-values calculated using robust standard errors at the residential prefecture level. Numbers without parentheses indicate elasticity, which is the standardized effect. Various control variables are included in all columns, such as residential prefecture and a dummy variable for marital status. Furthermore, the probabilities that the dependent variables have 2, 3, and 4 are calculated but not reported here.

The results are reported in Table 3 under the setting that GAI may have caused misinformation. The SMART results show a significantly positive sign in columns (1) and (2). Furthermore, its elasticity is approximately 0.90, which is sizable. Hence, smartphone usage time prevents individuals from caring about the possibility of misinformation. Conversely, COMPUTER shows a negative and statistically significant sign in columns (1) and (2). Thus, computer uage time allows individuals to avoid misinformation while creating intellectual documents. The results in Tables 2 and 3 together imply that the probability of mixing misinformation remarkably changes the computer user's views about GAI. Based on their experience using computers, individuals consider it important to cope with uncertainty. Meanwhile, smartphones enable users to write simple, short messages, whereas computers enable them to write complicated, long documents. This difference seems to result in the opposite signs. The statistical significance of INCOME disappears once we consider the possibility of misinformation.

FEMALE and AGE have significantly negative signs. Furthermore, the absolute values of elasticities are far larger than those in Table 1. In particular, the elasticity of FEMALE is approximately 0.17, which is 1.5 times larger than 0.11 in column (2) of Table 1. This is in line with female's risk-aversion tendencies [26–29]. Considering the possibility of misinformation causes TOP MANAGE and ORDINARY to have positive signs and statistical significance at the 1% level. Meanwhile, TEACHER has a negative sign and is statistically significant at the 1% level. Interestingly, the statistical significance of MEDICAL disappears. Compared to other industries and sectors, the medical sector is directly related to human life. Therefore, misinformation can even cause human deaths. Therefore, the existence of misinformation has changed medical practitioners' views on GAI.

Overall, we observe differences in the views of people once they are asked whether they care about misinformation being present in GAI's output.

## 4.3 Results on views about autonomous AI

**Table 4. Dependent variable: Views about autonomous AI (ordered logit model).**

Elasticity (standardized effect).

|  | (1)<br>*Typical Autonomous* | (2)<br>*Intellectual Autonomous* |
|---|---|---|
| **Prob (=5)** | | |
| UNIV | −0.035<br>(−1.15) | −0.028<br>(−0.98) |
| SMART | 0.096***<br>(3.55) | 0.104***<br>(5.10) |
| COMPUTER | 0.010<br>(0.24) | −0.035<br>(−0.70) |
| INCOME | 0.045<br>(1.09) | 0.042<br>(1.07) |
| FEMALE | −0.249***<br>(−8.97) | −0.251***<br>(−7.32) |
| AGE | −0.848***<br>(−6.51) | −1.024***<br>(−7.54) |
| TOP MANAGE | 0.019**<br>(2.07) | 0.012<br>(1.14) |
| ORDINARY | 0.078***<br>(3.56) | 0.060**<br>(2.57) |
| TEACHER | −0.001<br>(−0.37) | −0.002<br>(−0.65) |
| SPECIAL | −0.003<br>(−0.63) | 0.001<br>(0.31) |
| MEDICAL | 0.003<br>(1.13) | 0.002<br>(0.90) |
| **Prob (=1)** | | |
| UNIV | 0.029<br>(1.16) | 0.022<br>(0.98) |
| SMART | −0.083***<br>(−3.48) | −0.088***<br>(−5.03) |
| COMPUTER | −0.008<br>(−0.24) | 0.028<br>(0.70) |
| INCOME | −0.038<br>(−1.08) | −0.039<br>(−1.07) |
| FEMALE | 0.194***<br>(9.37) | 0.191***<br>(7.69) |
| AGE | 0.698*** | 0.827*** |
| | (6.55) | (7.62) |
| TOP MANAGE | −0.017** | −0.010 |
| | (−2.02) | (−1.13) |
| ORDINARY | −0.069*** | −0.053*** |
| | (−3.45) | (−2.50) |
| TEACHER | 0.001 | 0.002 |
| | (0.38) | (0.68) |
| SPECIAL | 0.002 | −0.001 |
| | (0.64) | (−0.31) |
| MEDICAL | −0.002 | −0.001 |
| | (−1.04) | (−0.84) |
| Pseudo $R^2$ | 0.02 | 0.03 |
| Observations | 3,433 | 3,433 |

Note: ***, **, and * denote statistical significance at the 1%, 5%, and 10% levels, respectively. The numbers within parentheses are the z-values calculated using robust standard errors at the residential prefecture level. Numbers without parentheses indicate elasticity, which is the standardized effect. Various control variables are included in all columns, such as residential prefecture and a dummy variable for marital status. Furthermore, the probabilities that the dependent variables have 2, 3, and 4 are calculated but are not reported.

The results in Table 4 are similar to those in Table 3. The effect of the probability of AI adding unrelated content is similar to that of the probability of AI adding misinformation. However, TOP MANAGE does not show statistical significance in column (2), suggesting that managers care about mixing undirect content in intellectual documents. The elasticity of FEMALE is approximately 2.0, which is larger than that shown in Table 2. Therefore, females are more likely to care about undirected content than about misinformation. The statistical significance of TEACHER disappears in columns (1) and (2). Therefore, teachers think more about misinformation than unintended content.

## 4.4 Results on views about AI replacing human

**Table 5. Dependent variable: View about AI replacing humans (ordered logit model).**

Elasticity (standardized effect).

|  | (1) *Self Typical* | (2) *self Intellectual* |
|---|---|---|
| **Prob (=5)** |  |  |
| UNIV | 0.023 | 0.003 |
|  | (0.68) | (0.07) |
| SMART | 0.046 | 0.079*** |
|  | (1.46) | (3.67) |
| COMPUTER | 0.146*** | 0.092** |
|  | (4.25) | (2.24) |
| INCOME | 0.243*** | 0.134*** |
|  | (4.26) | (2.86) |
| FEMALE | −0.049** | −0.148*** |
|  | (−2.00) | (−5.50) |
| AGE | −0.236 | −0.567*** |
|  | (−1.29) | (−3.01) |
| TOP MANAGE | −0.003 | 0.017* |
|  | (−0.44) | (1.96) |
| ORDINARY | 0.001 | 0.035** |
|  | (0.07) | (2.23) |
| TEACHER | −0.0001 | −0.008 |
|  | (−0.04) | (−1.60) |
| SPECIAL | −0.005 | 0.002 |
|  | (−1.25) | (0.32) |
| MEDICAL | 0.005*** | 0.003** |
|  | (3.17) | (1.98) |
| **Prob (=1)** |  |  |
| UNIV | −0.025 | −0.002 |
|  | (−0.68) | (−0.07) |
| SMART | −0.050 | −0.078*** |
|  | (−1.44) | (−3.61) |
| COMPUTER | −0.161*** | −0.094** |

|  | (−4.15) | (−2.21) |
|---|---|---|
| INCOME | −0.270*** | −0.133*** |
|  | (−4.12) | (−2.81) |
| FEMALE | 0.051** | 0.138*** |
|  | (2.02) | (5.68) |
| AGE | 0.252 | 0.547*** |
|  | (1.30) | (3.03) |
| TOP MANAGE | 0.003 | −0.017* |
|  | (0.44) | (−1.92) |
| ORDINARY | −0.002 | −0.036** |
|  | (−0.07) | (−2.20) |
| TEACHER | 0.0002 | 0.007* |
|  | (0.04) | (1.86) |
| SPECIAL | 0.006 | −0.002 |
|  | (1.30) | (−0.32) |
| MEDICAL | −0.007** | −0.004* |
|  | (−2.55) | (−1.86) |
| Pseudo $R^2$ | 0.01 | 0.02 |
| Observations | 3,433 | 3,433 |

Note: ***, **, and * denote statistical significance at the 1%, 5%, and 10% levels, respectively. The numbers within parentheses are the z-values calculated using robust standard errors at the residential prefecture. Numbers without parentheses indicate elasticity, which is the standardized effect. Various control variables are included in all columns, such as residential prefecture and a dummy variable for marital status. Furthermore, the probabilities that the dependent variables have 2, 3, and 4 are calculated but are not reported.

Considering their own situations may change people's view from their views for humans in general. Some results in columns (1) and (2) of Table 5 differ from those in columns (1) and (2) of Table 2. The positive sign of SMART becomes statistically significant for producing intellectual documents. Communication through smartphones is typically been done by short messages. This lack of skill naturally leads them to depend on GAI for intellectual documents but not for typical documents. FEMALE shows a negative sign and statistical significance in both columns, although it is not statistically significant in column (1) of Table 2. That is, females prefer producing any type of document for themselves and not depend on GAI, although they do not disagree with the view that other people generally use GAI to produce typical documents. Furthermore, its absolute value of elasticity is approximately 0.15 in column (2) in Table 5, which is larger than 0.119 in column (2) of Table 2.

Contrary to Frey and Osborne [4], SPECIAL is not statistically significant. That is, individuals in special occupations are not opposed to them being replaced by GAI. We observe differences between lawyers and accountants, even when they are categorized as SPECIAL. Lawyer perform various tasks, such as building trusting relationships with clients and gathering various pieces of evidence advantageous for clients through field research. Understandably, these tasks cannot be conducted by AI. Rather, AI is useful as a tool to support human's tasks for some special workers, such as lawyers. The significantly positive sign of MEDICAL may reflect that the practical applications of AI are advancing in diagnostic imaging and testing. This allows workers to allocate their time and energy to the work that they are better at than those by AI. Thus, AI is replacing part of the work that humans perform, thereby improving their efficiency and reducing their overwork. Hence, medical workers agree that AI may replace their tasks.

## 4.5 Results on views about the role of AI

**Table 6. Dependent variable: Views about the role of AI (ordered logit model).**

Elasticity (standardized effect).

|  | (1) *Confidential* | (2) *Human Decision* | (3) *GAI Decision* |
|---|---|---|---|
| **Prob (=5)** | | | |
| UNIV | −0.010 | −0.020 | −0.032 |
|  | (−0.36) | (−0.59) | (−0.93) |
| SMART | 0.047 | 0.040 | 0.047** |
|  | (1.65) | (1.62) | (1.99) |
| COMPUTER | −0.095*** | 0.088*** | 0.022 |
|  | (−2.68) | (3.34) | (0.59) |
| INCOME | −0.083 | 0.011 | −0.015 |
|  | (−1.12) | (0.23) | (−0.26) |
| FEMALE | −0.163*** | −0.210*** | −0.236*** |
|  | (−6.57) | (−10.44) | (−9.85) |
| AGE | −1.159*** | −0.900*** | −1.147*** |
|  | (−7.14) | (−4.78) | (−6.12) |
| TOP MANAGE | 0.037*** | 0.012 | 0.010 |
|  | (3.80) | (1.22) | (1.14) |
| ORDINARY | 0.100*** | 0.080*** | 0.072*** |
|  | (5.42) | (3.44) | (4.49) |
| TEACHER | −0.012*** | −0.013*** | −0.009*** |
|  | (−3.22) | (−3.30) | (−2.69) |
| SPECIAL | −0.002 | −0.003 | −0.001 |
|  | (−0.48) | (−0.43) | (−0.27) |
| MEDICAL | 0.001 | 0.001 | 0.003 |
|  | (0.38) | (0.72) | (1.62) |
| **Prob (=1)** | | | |
| UNIV | 0.007 | 0.017 | 0.025 |
|  | (0.36) | (0.59) | (0.93) |
| SMART | −0.034 | −0.034 | −0.039** |
|  | (−1.63) | (−1.60) | (−1.98) |
| COMPUTER | 0.066*** | −0.076*** | −0.018 |
|  | (2.73) | (−3.30) | (−0.59) |
| INCOME | 0.058 | −0.009 | 0.012 |
|  | (1.12) | (−0.23) | (0.26) |
| FEMALE | 0.107*** | 0.169*** | 0.178*** |
|  | (6.95) | (11.20) | (10.24) |
| AGE | 0.794*** | 0.757*** | 0.908*** |
|  | (7.22) | (4.78) | (6.20) |
| TOP MANAGE | −0.028*** | −0.011 | −0.009 |

|  | (−3.53) | (−1.21) | (−1.12) |
|---|---|---|---|
| ORDINARY | −0.078*** | −0.072*** | −0.063*** |
|  | (−5.15) | (−3.37) | (−4.38) |
| TEACHER | 0.005*** | 0.009*** | 0.006*** |
|  | (5.59) | (4.60) | (3.38) |
| SPECIAL | 0.001 | 0.003 | 0.001 |
|  | (0.4) | (0.44) | (0.27) |
| MEDICAL | −0.001 | −0.001 | −0.003 |
|  | (−0.37) | (−0.70) | (−1.50) |
| Pseudo $R^2$ | 0.03 | 0.02 | 0.03 |
| Observations | 3,433 | 3,433 | 3,433 |

Note: ***, **, and * denote statistical significance at the 1%, 5%, and 10% levels, respectively. The numbers within parentheses are the z-values calculated using robust standard errors at the residential prefecture level. Numbers without parentheses indicate elasticity, which is the standardized effect. Various control variables are included in all columns, such as residential prefecture and a dummy variable for marital status. Furthermore, the probabilities that the dependent variables have 2, 3, and 4 are calculated but are not reported here.

Next, we examine issues profoundly related to whether humans will play a role in business and society because of GAI development. Interestingly, the view that GAI can produce confidential documents differs between smartphone and computer usage times. Column (1) shows a positive sign for SMART and a significantly negative sign for COMPUTER. The absolute value of COMPUTER is 0.086, which is sizable. Using a computer prevents individuals from using GAI to create confidential documents. This may be because computers can be used for more confidential work than smartphones. Naturally, through the experience of dealing with confidential and secret information, individuals become aware of the dangers of using AI related to confidential information leakage.

Regarding human decision-making, we compare the results in columns (2) and (3). COMPUTER show a significantly positive sign for human decision making. Meanwhile, the statistical significance of COMPUTER disappears for the view that it is better for GAI to make decisions on management and other matters. Using a computer allows individuals to recognize the limitations of using GAI. Thus, the more proficient individuals become with computers, the more they know which areas humans are better at than GAI. Therefore, they consider the division of labor between GAI and humans to be important. GAI is a useful tool for improving efficiency by replacing humans, although humans allocate their time and energy to what they can do better than GAI.

FEMALE and AGE continue to have signficant negative signs. The absolute values of elasticity are approximately 0.2 and 1.0, respectively. These values are approximately three to four times larger than those in Tables 2-5. Clearly, females and older persons are opposed to a reduction in human roles. They seemingly do not trust GAI to replace humans in various critical tasks previously conducted by humans, such as writing confidential documents and making decisions.

Both TOP MANAGE and ORDINARY show significant positive signs for the views regarding using GAI in creating confidential documents in column (1). Surprisingly, managers are considering replacing humans with GAI for confidential works. One possible interpretation is that they have underestimated the risk of leakage through GAI. Otherwise, they consider the confidentiality risks related to human to be equivalent to those with GAI.

Turning to decision making, we differentiate between TOP MANAGE and ORDINARY in columns (2) and (3). The significant positive sign of ORDINARY not only supports the view that humans only need to make final decisions, but also the view that GAI makes decisions on management and other tasks. That is, ordinary employees believe that humans will no longer work in the future. However, TOP MANAGE is not statistically significant for these views. Through their experience in managerial tasks, they have a bird's eye view of the corporate organization and the business world, and are familiar with the tasks and skills that cannot be replaced by GAI for the business. Accordingly, they know that GAI simply cannot replace humans in performing various tasks.

SPECIAL shows a significant sign, which is consistent with the results in Table 5. The statistical insignificance of MEDICAL implies that humans continue to perform tasks better than GAI. Using GAI is effective for supporting clinical and healthcare decision-making [30,31]. Considering the results for MEDICAL in Tables 2-6 together suggests that medical workers consider the division of labor between human labor and GAI to be significant, although some human tasks should be replaced by GAI.

## 4.6 Results on views about the importance of human learning

**Table 7. Dependent variable: Views about the importance of human learning (ordered logit model).** Elasticity (standardized effect).

|  | (1) *No Writing* | (2) *No Foreign* | (3) *No Learning* |
|---|---|---|---|
| **Prob (=5)** | | | |
| UNIV | −0.017 | −0.033 | −0.028 |
|  | (−0.47) | (−0.74) | (−0.78) |
| SMART | 0.060** | −0.001 | 0.072*** |
|  | (2.55) | (−0.03) | (3.01) |
| COMPUTER | −0.094*** | 0.053 | −0.104*** |
|  | (−3.00) | (1.54) | (−3.18) |
| INCOME | −0.101 | −0.035 | −0.134* |
|  | (−1.38) | (−0.69) | (−1.72) |
| FEMALE | −0.183*** | −0.135*** | −0.204*** |
|  | (−5.95) | (−3.94) | (−6.44) |
| AGE | −1.304*** | −1.073*** | −1.393*** |
|  | (−8.66) | (−6.12) | (−8.41) |
| TOP MANAGE | 0.034*** | 0.019 | 0.034*** |
|  | (2.72) | (1.61) | (3.04) |
| ORDINARY | 0.108*** | 0.079*** | 0.115*** |
|  | (5.35) | (3.74) | (7.13) |
| TEACHER | −0.013** | −0.011*** | −0.011*** |
|  | (−2.55) | (−2.90) | (−2.63) |
| SPECIAL | −0.001 | −0.003 | 0.001 |
|  | (−0.19) | (−0.53) | (0.15) |
| MEDICAL | −0.001 | 0.004* | −0.001 |
|  | (−0.31) | (1.82) | (−0.35) |
| **Prob (=1)** | | | |
| UNIV | 0.012 | 0.027 | 0.018 |
|  | (0.47) | (0.74) | (0.79) |
| SMART | −0.045** | 0.001 | −0.050*** |
|  | (−2.50) | (0.03) | (−2.96) |
| COMPUTER | 0.065*** | −0.044 | 0.067*** |
|  | (3.05) | (−1.53) | (3.25) |
| INCOME | 0.071 | 0.028 | 0.087* |
|  | (1.39) | (0.69) | (1.74) |
| FEMALE | 0.120*** | 0.106*** | 0.122*** |
| | (6.30) | (4.11) | (6.97) |
| AGE | 0.896*** | 0.871*** | 0.889*** |
| | (8.75) | (6.21) | (8.69) |
| TOP MANAGE | −0.026*** | −0.017 | −0.024*** |
| | (−2.58) | (−1.57) | (−2.89) |
| ORDINARY | −0.085*** | −0.070*** | −0.087*** |
| | (−5.14) | (−3.64) | (−6.75) |
| TEACHER | 0.006*** | 0.007*** | 0.005*** |
| | (4.67) | (4.20) | (4.43) |
| SPECIAL | 0.001 | 0.002 | −0.001 |
| | (0.19) | (0.54) | (−0.15) |
| MEDICAL | 0.0004 | −0.004 | 0.005 |
| | (0.31) | (−1.62) | (0.36) |
| Pseudo $R^2$ | 0.02 | 0.02 | 0.03 |
| Observations | 3,433 | 3,433 | 3,433 |

Note: ***, **, and * denote statistical significance at the 1%, 5%, and 10% levels, respectively. The numbers within parentheses are the z-values calculated using robust standard errors at the residential prefecture level. Numbers without parentheses indicate elasticity, which is the standardized effect. Various control variables are included in all columns, such as residential prefecture and a dummy variable for marital status. Furthermore, the probabilities that the dependent variables have 2, 3, and 4 are calculated but are not reported here.

Human learning would not be needed if GAI generally outperforms a skilled and intelligent human. Then, allocating human resources, such as time and energy, from learning to other tasks can improve economic efficiency. Table 6 reports the results regarding the views on humans learning writing, humans learning foreign languages, and general human learning.

Except learning a foreign language, smartphone usage time is significantly positively associated with the opinion that human learning is unnecessary. Meanwhile, computer usage time shows a significant negative correlation with the opinion that human learning is unnecessary. This may because smartphone users consider GAI as a substitute for human learning, whereas computer users regard GAI as a complement.

By using GAI, businesspeople can save resources, such as time and energy, for learning. Hence, GAI is preferred to human learning if the outcome is the same, thereby improving efficiency. Surprisingly, TOP MANAGE and ORDINARY exhibit significant positive signs. In most cases, both managers and ordinary workers agree with the view of unnecessary human learning. However, unlike ordinary employees, managers do not have a positive view of foreign-language learning being unnecessary. This may be because GAI does not enable managers to have good human communication and business negotiations cannot be successful using GAI.

The significant negative sign of TEACHER is consistent with the intuition that teachers consider human learning to be necessary, which gives them the opportunity to become teachers. However, another interpretation is reasonable: it may be possible that reducing the demand for learning will lead teachers to lose their jobs. However, this hypothesis is supported by the results in Table 5, which show that TEACHER is not statistically significant. The results for FEMALE and AGE are similar to those in Table 6. Therefore, these respondents consider it essential for humans to perform critical tasks, learn to improve skills, and enhance their comprehensive human capabilities.

Neither SPECIAL nor MEDICAL have statistical significance in most cases. To become professional workers in these areas, people are generally required to learn and acquire skills to pass certification examinations. Furthermore, humans are necessary for these tasks, at least partially, because some of the tasks cannot be replaced by AI. Therefore, these workers consider learning to be important.

# 5 Discussion

## 5.1 Theoretical and practical contributions

This study makes some theoretical contributions. An increasing number of studies regarding GAI deal with efficiency and decision-making in business [10,32–35], and effectiveness as an educational tool [13,14,16,36,37] and a support for medical work [11,30,31,38]. Meanwhile, to the best of our knowledge, this study is the first to comprehensively assess how individuals' characteristics are associated with views about GAI in considering various settings. We contribute to the literature by revealing the subjective views about not only GAI producing documents, but also GAI's influence on human roles and human learning.

This study also has some practical implications. Having experience of using smartphone leads individuals to disregard human learning, whereas the computer usage experience leads them to focus on human learning. Assuming that human learning generates human's creativity, excessive smartphone usage may hamper creativity. Next, ordinary employees agree that GAI can replace humans not only in routine work but also in decision-making. However, managers do not agree with this. Rather, managers believe that humans should perform various tasks and make decisions, whereas GAI can replace humans to produce documents. Surprisingly, neither ordinary employees nor managers place importance on human learning if the GAI develops sufficiently. That is, in the business world, creativity through human learning is not necessary because only typical tasks are performed.

Next, the data set is gathered from Japanese individuals. Intuitively, Japanese managers' strategies are more likely to be common in Japan than in other countries, such as the US [39]. In the present case, creativity and learning are not required in businesses. Businesspersons expect GAI not only to function as a tool to reduce unproductive work, such as producing documents, but also to be useful in reducing the time spent on learning. However, besides producing documents, various other tasks are necessary to improve firm performance. Managers do not consider GAI to be effective in forming human networks, coordinating teams, and encouraging and increasing employee motivation.

Finally, females generally have negative views of GAI, presumably because using GAI is risky in the real world. This is consistent with studies showing that females are more risk averse than males [26–29]. The meaning and significance of learning and working is not only to improve productivity, but also to improve subjective well-being. In other words, learning and work are thought to enrich human life and promote health. Females should consider these points more carefully. This widening perspective seems to have changed the view of the role of the GAI.

## 5.2 Limitations

People's views about GAI in education can vary by country. Given that the cultural and educational context in Japan is different from that in other countries, our findings based on Japanese respondents may differ from those from other countries [14]. Additional research should be conducted in countries with different socioeconomic backgrounds. Furthermore, this study's findings were based on a dataset gathered in February 2024. However, GAI has and will continue to develop and improve rapidly. Additionally, the number of GAI users is rapidly increasing. Naturally, the views about GAI are likely to change; therefore, additional research using panel data is required.

# 6  Conclusion

Since the end of 2022, GAI has been developing and modifying rapidly in the public domain. Crucially, people's views about it have changed during its development process. Our survey showed that at the beginning of 2024, views about GAI varied widely between the types of work, gender, and time spent on smartphones and computers.

GAI can perform unproductive tasks, such as writing documents. Especially in the business world, GAI is accepted as a tool to reduce unproductive work done by human labor, and thus, improve economic efficiency. Hence, the time spent on manual tasks is likely to decrease for humans, possibly increasing leisure time to enjoy life more. Will GAI make things more rosy?

Surprisingly, some groups supported the view that GAI can replace humans in all tasks, including decision-making, and that human learning is unnecessary if GAI develops sufficiently. Human learning leads humans to think deeply; that is, "I think, therefore I am." Can GAI render thinking irrelevant for humans? Clearly, GAI raises questions regarding the meaning of human existence.

# Acknowledgments

We thank Editage (http://www.editage.com) for English language editing and review.

## Data Availability.

The datasets analysed during the current study are available in the RESEACH MAP repository as follows;

https://researchmap.jp/multidatabases/multidatabase_contents/detail/712923/5d6ae7a9b98e14c68aed254d83715577?frame_id=1322935

5.  Larsen, B. G. AI: A game-changer society needs to be ready for. *World Economic Forum* (2023).

6.  Stokel-Walker, C. ChatGPT listed as author on research papers: many scientists disapprove. *Nature* **613**, 620–621 (2023).

7.  Pavlik, J. V. Collaborating With ChatGPT: Considering the Implications of Generative Artificial Intelligence for Journalism and Media Education. *Journalism & Mass Communication Educator* **78**, 84–93 (2023).

8.  Hwang, T. *et al.* Can ChatGPT assist authors with abstract writing in medical journals? Evaluating the quality of scientific abstracts generated by ChatGPT and original abstracts. *PLoS One* **19**, (2024).

9.  Noy, S. & Zhang, W. Experimental evidence on the productivity effects of generative artificial intelligence. *Science (1979)* **381**, 187–192 (2023).

10. Chuma, E. L. & De Oliveira, G. G. Generative AI for Business Decision-Making: A Case of ChatGPT. *Management Science and Business Decisions* **3**, 5–11 (2023).

11. Hosseini, M. *et al.* An exploratory survey about using ChatGPT in education, healthcare, and research. *PLoS One* **18**, (2023).

12. Terwiesch, C. *Would Chat GPT3 Get a Wharton MBA? A Prediction Based on Its Performance in the Operations Management Course*. https://mackinstitute.wharton.upenn.edu/wp-content/uploads/2023/01/Christian-Terwiesch-Chat-GTP.pdf (2023) doi:doi.org/10.2139/ssrn.4380365.

13. Yu, H. Reflection on whether Chat GPT should be banned by academia from the perspective of education and teaching. *Frontiers in Psychology* vol. 14 Preprint at https://doi.org/10.3389/fpsyg.2023.1181712 (2023).

14. Ibrahim, H. *et al.* Perception, performance, and detectability of conversational artificial intelligence across 32 university courses. *Sci Rep* **13**, (2023).

15. Kalla, D., Smith, N., Samaah, F. & Kuraku, S. Study and Analysis of Chat GPT and its Impact on Different Fields of Study. *International Journal of Innovative*
43

*and Technology* **3**, 60–68 (2023).

37. Herbold, S., Hautli-Janisz, A., Heuer, U., Kikteva, Z. & Trautsch, A. A large-scale comparison of human-written versus ChatGPT-generated essays. *Sci Rep* **13**, (2023).

38. Hulman, A. *et al.* ChatGPT- versus human-generated answers to frequently asked questions about diabetes: A Turing test-inspired survey among employees of a Danish diabetes center. *PLoS One* **18**, (2023).

39. Ohtake, F. & Ohkusa, Y. Testing the Matching Hypothesis: The Case of Professional Baseball in Japan with Comparisons to the United States. *J Jpn Int Econ* **8**, 204–219 (1994).